# Ring statistics of silica bilayers

Avishek Kumar[1], David Sherrington[2], Mark Wilson[3] and M. F. Thorpe*[1,2]

[1] Physics Department, Arizona State University, Tempe, AZ 85287, USA
[2] Rudolf Peierls Centre for Theoretical Physics, University of Oxford, 1 Keble Rd, Oxford OX1 3NP.
[3] Department of Chemistry, Physical and Theoretical Chemistry Laboratory, University of Oxford, South Parks Road, Oxford OX1 3QZ, England

*Corresponding author: e-mail mft@asu.edu

**Abstract** The recent synthesis and characterisation of bilayers of vitreous silica has produced valuable new information on ring sizes and distributions. In this paper, we compare the ring statistics of experimental samples with computer generated samples. The average ring size is fixed at six by topology, but the width, skewness and other moments of the distribution of ring edges are characteristics of particular samples. We examine the Aboav-Weaire law that quantifies the propensity of smaller rings to be adjacent to larger rings, and find similar results for available experimental samples which however differ somewhat from computer-generated bilayers currently. We introduce a new law for the areas of rings of various sizes.

**1 Introduction.** The 80 years since Zachariasen's famous paper [1] on the random network theory of glass structure have seen steady progress in our understanding of the structure of glassy materials through the construction of models and comparison with experiment.[2,3] In the early days, models were hand-built with plastic units and had free boundary conditions.[4] Today, very much larger computer models have periodic boundary conditions.[5-7] Although the pair distribution function can be found by Fourier transforming experimental diffraction data, this is a relatively imprecise tool, and is largely insensitive to the details of the intermediate-range order. This has been frustrating as the range of vitreous silica structures, characterised by the fictive temperature, most likely differ in the details of the connectivities of the local tetrahedral coordination polyhedral as probed by the ring statistics [8]. Other experimental observations, such as the $D_1$ and $D_2$ lines observed in the Raman spectra, are thought to be directly linked to localised modes on specific rings [9], but this remains to be confirmed by direct experimental evidence.

Recently, two experimental groups [10, 11] have synthesised and imaged samples that consist of bilayers of vitreous silica. This has provided new insight into the structure of glass; *albeit a two-dimensional glass*. In Figure 1, we show a computer model of the bilayer structure from various perspectives. The lower panel shows a side view of the bilayer. It has been argued by us [12] that there is a mirror plane involving the central oxygen atoms, shown as red dots in the lower panel of Figure 1, as well as a slight puckering of the upper and lower oxygen surfaces of the bilayer, consistent with maintaining the symmetry. The middle panel of Figure 1 shows the projection of a single monolayer onto the plane, with the projection of the silicon atoms lying at the centers of the yellow triangles. The top panel shows the network of silicon atoms deduced from the middle panel and it is this representation that we will focus on in this note. The reasons for this are two-fold. The first is that this is a convenient minimalist representation of the network topology that contains all the important information about ring statistics, as well as the Aboav-Weaire [13, 14] and ring areas The second reason is that there are other examples of similar networks, where every vertex has edges to three other vertices; examples are soap films, the Giant's Causeway (in Northern Ireland), biological tissue and many others [15, 16]. Much effort has been used to characterise and compare such networks and we will use similar methods here.

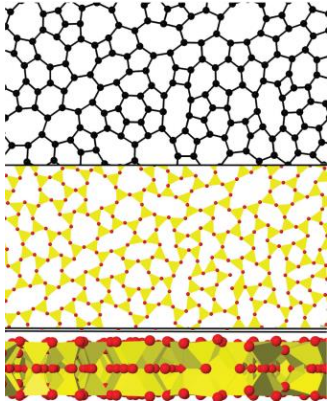

**Figure 1** *The lower panel shows a side view of the silica bilayer. The central panel shows the projection in the plane of one monolayer with the red dots representing oxygen atoms, and with a silicon atom in the middle of each yellow triangle. The upper panel is dual to the central panel and shows the topology of the silicon network, and is the focus of this paper. Adapted from Figure 2 of reference [12].*

Although the networks studied here are two-dimensional and so not representative of bulk three-dimensional glasses, they are important in their own right and for the insight they give into bulk glasses. The advantage of having two dimensional random networks cannot be over-emphasized; not least as visualisation is so much easier than for three dimensional networks.

**2 Ring statistics.** We have collected together a number of experimental results [10, 11] and some theoretical computer modelling results [12] for amorphous silica bilayers to examine some simple measures of the ring statistics. All networks have been computer-refined to produce networks of perfect tetrahedra (with all six edges having equal length) that share corners. Their two-dimensional projections are shown in Figure 2. The edges, which represent the Si-Si separations, do not all have equal lengths, partially because of the projection into the plane, but mainly because of the variation in the Si-O-Si angle [12]. Note that the Si-O bond length is fixed by the covalent chemistry at ~1.6Å and is effectively independent of the local environment.

One could question how truly 'random' these networks are; visual inspection suggests some regions of yellow sixfold rings that are somewhat larger than statistically would be expected, as for example in the left side of sample (e). This is an important question for future study. However, here we will simply focus on the ring statistics as given in Table 1, which can be obtained by inspection from Figure 2 and Figure 3.

|  | $p_4$ | $p_5$ | $p_6$ | $p_7$ | $p_8$ | $p_9$ | $p_{10}$ | $\langle r \rangle$ | $\mu_2$ | $\gamma_1$ | $-\gamma_2$ | α |
|---|---|---|---|---|---|---|---|---|---|---|---|---|
| Exp. sample (a) | 0.04 | 0.27 | 0.42 | 0.21 | 0.05 | 0.01 | 0.002 | 6.00 | 0.94 | 0.39 | 1.89 | 0.33 |
| Exp. sample (b) | 0.04 | 0.28 | 0.45 | 0.16 | 0.07 | 0.01 | 0 | 5.97 | 0.94 | 0.42 | 1.89 | 0.36 |
| Exp. sample (c) | 0.04 | 0.29 | 0.47 | 0.10 | 0.10 | 0 | 0 | 5.94 | 0.95 | 0.50 | 1.88 | 0.31 |
| Exp. sample (d) | 0.08 | 0.25 | 0.42 | 0.20 | 0.03 | 0.03 | 0 | 5.95 | 1.15 | 0.57 | 1.87 | 0.36 |
| Exp. sample (e) | 0.03 | 0.26 | 0.45 | 0.20 | 0.04 | 0.004 | 0.002 | 5.98 | 0.84 | 0.33 | 1.90 | 0.30 |
|  |  |  |  |  |  |  |  |  |  |  |  |  |
| Comp. sample (f) | 0.02 | 0.33 | 0.37 | 0.21 | 0.07 | 0.005 | 0 | 6 | 0.94 | 0.42 | 1.89 | 0.18 |
| Comp. sample (g) | 0 | 0.37 | 0.32 | 0.25 | 0.06 | 0 | 0 | 6 | 0.86 | 0.36 | 1.90 | 0.23 |

**Table 1** *Showing the ring statistics via the probability $p_r$, the mean ring size $\langle r \rangle$, the second moment $\mu_2$, the skewness $\gamma_1$ and the excess kurtosis $\gamma_2$ of the ring distribution for the five experimental samples shown in Figure 2 and the two computer generated samples shown in Figure 3. In the last column, the values of the parameter α are given that are used in the Aboav-Weaire law fits in Figures 5 and 6.*

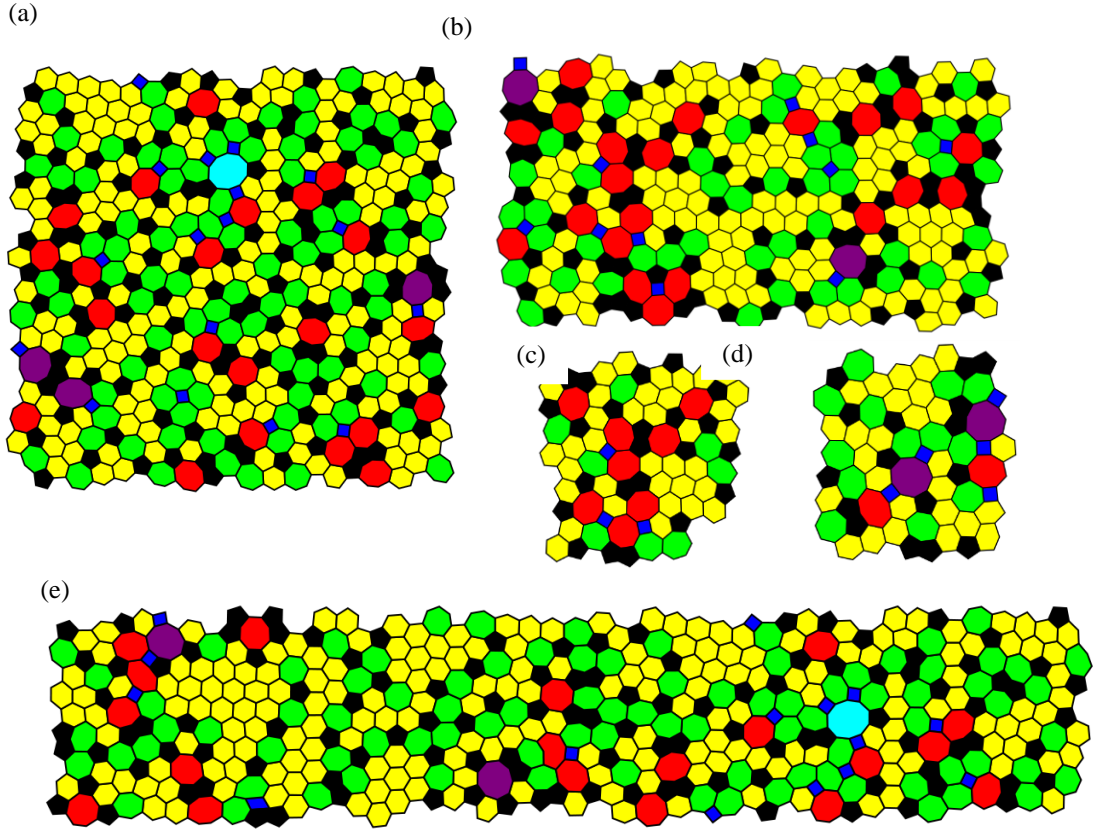

**Figure 2** *The ring structures obtained from five experimental samples (a) Cornell A [11], (b) Berlin C [10], (c) Berlin A [10], (d) Berlin B [10] and (e) Cornell B [11]. The ring colouring is blue (4), black (5), yellow (6), green (7), red (8), purple (9) and pale blue (10), where the number in brackets is the ring size.*

The distribution of ring sizes, or local ring statistics, of these networks is shown in Table 1, where we denote the probability of having an $r$-sided ring as $p_r$, normalised so that

$$\sum_r p_r = 1. \tag{1}$$

The moments of the distribution $\langle r^s \rangle$ are given by

$$\langle r^s \rangle = \sum_r r^s \, p_r, \tag{2}$$

where for an infinite system $\langle r \rangle = 6$ from Euler's theorem [13 - 16]. It is convenient to characterise the width of the distribution using the second moment of the deviation from the mean value $\langle r \rangle = 6$

$$\mu_2 = \langle (r - \langle r \rangle)^2 \rangle = \langle r^2 \rangle - \langle r \rangle^2, \tag{3}$$

the skewness $\gamma_1$

$$\gamma_1 = \langle (r - \langle r \rangle)^3 \rangle, \tag{4}$$

and the excess kurtosis $\gamma_2$

$$\gamma_2 = \frac{\langle(r-\langle r\rangle)^4\rangle}{\langle(r-\langle r\rangle)^2\rangle^2} - 3. \qquad (5)$$

There is close agreement for the ring statistics among the experimental samples, even though the two groups used different preparation procedures, with the six-fold rings being the most probable at around 45%, with significant numbers of five and seven-fold rings at probabilities over 20%. There is most variability between samples in the probability of both small (4) and large rings (8, 9 and 10) as would be expected with so few of such rings and hence limited statistics. As a result the second moment $\mu_2$ and the skewness $\gamma$ also do not vary much among the five experimental samples.

In Figure 3, we show the two computer-generated samples (f) and (g) (obtained as described in ref. [12]) that have the closest second moment $\mu_2$ to the experimental results, and Table 1 lists the associated statistics. However, it can be seen that these computer generated samples show fewer sixfold rings and more fivefold rings than the experimental samples, and hence a larger skewness. More computer modelling needs to be performed to reproduce more precisely these experimentally observed ring statistics.

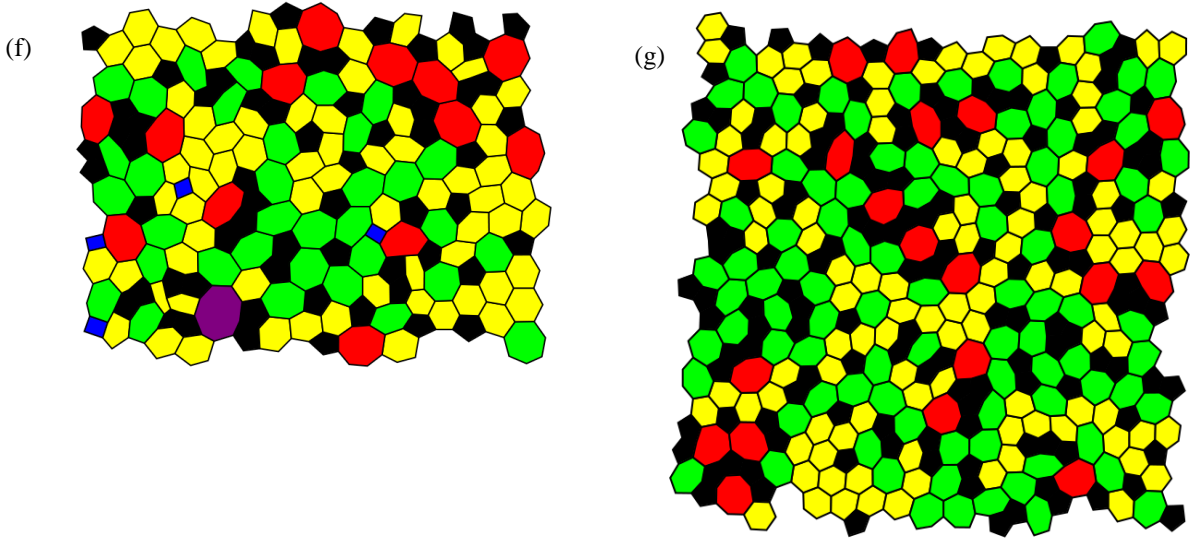

Figure 3 *Two computer generated samples, [12] (f) and (g) (referred to as 432(b) 836 respectively in (12)). Here the numbers correspond to the silicon atoms in each monolayer in each supercell. The samples have periodic boundary conditions and the figure here shows most of a single supercell. The ring colouring is again blue (4), black (5), yellow (6), green (7), red (8), purple (9) and pale blue (10), where the number in brackets is the ring size.*

In Figure 4, we show the cumulative ring statistics for all five of the experimental samples shown in Figure 2. This provides the best estimate of ring statistics from currently available experimental samples with $p_4 = 0.038$, $p_5 = 0.27$, $p_6 = 0.44$, $p_7 = 0.19$, $p_8 = 0.054$, $p_9 = 0.0075$ and $p_{10} = 0.0015$.

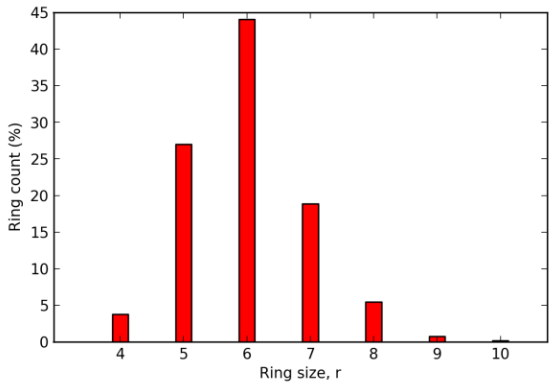

**Figure 4** *Showing the cumulative ring statistics for the five experimental samples (a) – (e) weighted by the size of the various samples. The vertical height of each red bar gives the percentage of rings of that size.*

**2 Aboav-Weaire law.** Non-crystalline glassy networks have historically been referred to as Continuous Random Networks (CRNs) [1-3] and this nomenclature is commonly used to characterise networks like those shown in Figures 1-3. In order that the mean ring size be 6, as required by Euler's theorem, and all the bonds of similar length, it is necessary for smaller rings to have a tendency to have larger rings as neighbours, sharing an edge, and vice versa. A popular way to characterise this propensity is through the mean size of the rings that surround a typical $r$-fold ring, $m_r$. A sum rule due to Weaire [14] gives

$$\sum_r m_r r p_r = 36 + \mu_2. \qquad (6)$$

Aboav [13] argued for the approximate rule for $m_r$

$$m_r = A + \frac{B}{r}. \qquad (7)$$

If we insert (7) into (6), we find that $6A = B = 36 + \mu_2$ and hence the Aboav-Weaire law can be written as

$$r m_r = (36 + \mu_2) + 6(1 - \alpha)(n - 6), \qquad (8)$$

which contains a single fitting parameter $\alpha$ [17]. Figures 5 and 6 show results for the five experimental networks of Figure 2 and the two computer-generated networks of Figure 3, plotted as $rm_r - \mu_2$ versus $r$ [*c.f.* Eq. (8)].

It can be seen that the fit to the Aboav-Weaire law is reasonably good, although certainly not exact, which there is no underlying mathematical reason to expect. We emphasize that is the relation (7) is not exact for general networks. The points at ring size $r = 6$, are close to the value of 36 shown by the horizontal yellow line. The variations between the five experimental samples are seen to be quite small and within the bounds expected for variations due to finite size effects. This is also shown by the values of $\alpha$ given in Table 1 which are all close to 1/3. A similar plot in Figure 6, for the two computer generated networks, shows similar results although with a slightly larger slope and hence smaller values of $\alpha$ given in Table 1.

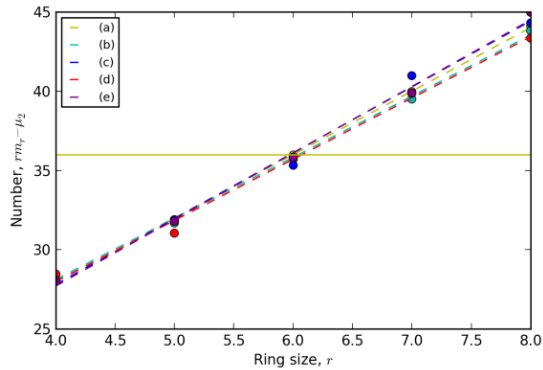

**Figure 5** *Plot of the Aboav-Weaire law for experimental samples as shown in the top left hand corner where the notation is from the caption to Figure 2. The yellow horizontal line at 36 shows the common value for ring size 6, from Eq. (8), and the values of alpha are given in the last column of Table 1.*

It can be seen that the fit to the various samples with Eq. (8) is not unreasonable although the statistics for high and low $r$ are not really adequate and much larger samples with many 10-fold rings etc would be needed for a more precise determination. However it is surprising that the subtle differences between the various networks is buried in the noise due to finite size effects, so there is less sensitivity than in the ring statistics $p_r$, which are the most obvious, and most important, quantities to focus on when comparing different networks. Clearly the ring-ring correlations are important and there is a need for new statistical approaches here, to better characterise and understand these networks. The computer generated samples do need further refinement, as can be seen by the lower values of the fraction of sixfold rings, $p_6$, and the smaller values of $\alpha$ when compared to experiment. These two deficiencies in the presently available computer generated models may well be related.

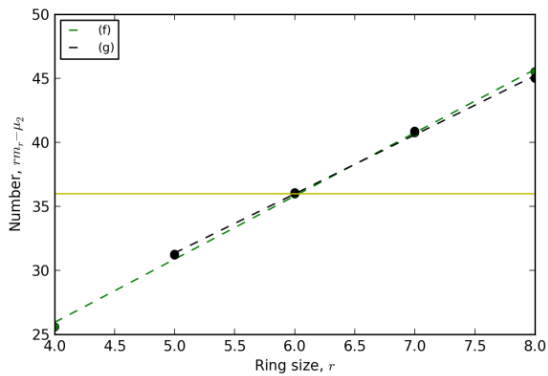

**Figure 6** *Plot of the Aboav-Weaire law for computer generated samples as shown in the top left hand corner where the notation is from the caption to Figure 3. The yellow horizontal line at 36 shows the common value for ring size 6, from Eq. (8), and the values of alpha are given in the last column of Table 1.*

**2 Area law.** Properties studied so far have been topological, in that no distance metric has been involved. Here we extend the analysis to include geometry. If the area of $r$-sided ring is $A_r$, averaged over the sample, then we write a dimensionless area $a_r$ as

$$a_r = \frac{A_r}{\langle l \rangle^2} \tag{9}$$

where $\langle l \rangle$ is the mean plane-projected length of an edge connecting pairs of vertices (i.e. silicon atoms) averaged over the whole sample. The term $\langle l \rangle^2$ in the denominator is to make the area $a_r$ dimensionless. Because of the nature of the bonding and the small tilt angle, there is only a variation of about $\pm 2\%$ in the lengths $l$ across the sample.

The area of a regular $r$-sided polygon with sides all of unit length is

$$a_r = \left[\frac{r}{4\tan(\pi/r)}\right], \tag{10}$$

which is plotted in Figures 7 and 8 as the solid blue line. The experimental points fit rather closely to this blue line (Figure 7), whereas the computer modelled points lie somewhat below (Figure 8). This demonstrates that the experimental samples have rings that are remarkably symmetric, with the maximum area possible, whereas the computer generated samples are less so with distortions that lower the area for most rings. More modelling work needs to be done to understand this discrepancy.

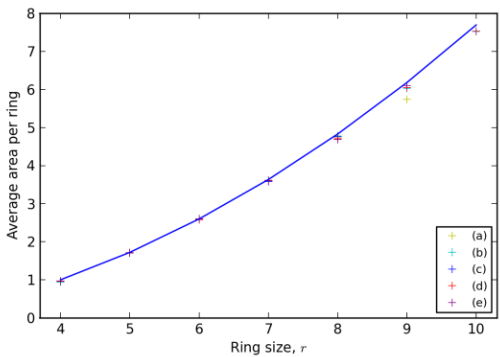

Figure 7 *The symbols show the average areas for rings of various sizes for experimental samples as shown in the bottom right hand corner where the notation is from the caption to Figure 2. The solid blue line is Eq. (10) with no adjustable parameters.*

Lewis [17] has studied similar, mainly biological networks, whose driving mechanisms are surely different from those of the essentially fixed-length bonds in vitreous silica bilayers. Also different in origin are networks in foams where the driving mechanism is surface tension. Lewis found that the area of the polygons increased linearly with ring size and this has come to be known as Lewis's law [17], which by visual inspection, clearly does not hold for the data here, as plotted in Figures 7 and 8..

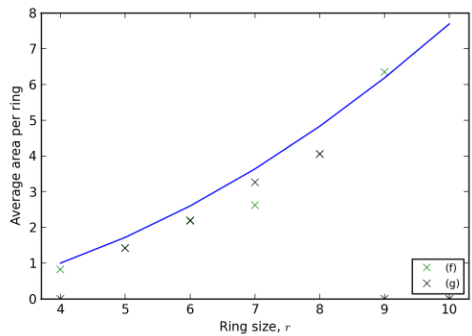

**Figure 8** *The symbols show the average areas for rings of various sizes for computer generated samples as shown in the bottom right hand corner where the notation is from the caption to Figure 3. The solid blue line is Eq. (10) with no adjustable parameters.*

**4. Discussion.** In this paper we have examined several properties of the ring statistics of two-dimensional amorphous networks with three-coordinated sites. These are derived from observed experimental results in silica bilayers in two separate studies using different techniques together with those obtained in corresponding computer modelling. These networks have a qualitative resemblance to the classic images of Zachariasen [1], as do two-dimensional foams, cuts through cells and natural phenomena such as Giant's Causeway [13-16]. However, these first atomic-level results have a very different bond and force character at a microscopic level, being constructed from units with highly constrained bond-lengths and with coordination conservation, in contrast to the surface tension driven forces of foams and the non-conservation of vertex number inherent in foams as a temporal reduction due to T2 processes and as an increasing quantity in cells through cell division processes [15-17]. Thus they provide a new system for the investigation of classic empirical *laws* such as those of Aboav-Weaire [13-14] and Lewis [15].

All the experimental silica samples measured to date have very similar ring statistics as measured for example by their second moments of the distribution of their cell-edge sizes. The computer generated models available to date are also similarly restricted. Although limited, this study already it shows interesting results, both of similarity and of difference compared with (the non atomic level) systems studied earlier in other manifestations. In particular, the Aboav-Weaire law appears to be reasonable, whereas Lewis' 'law' is not; a simple regular-polygon idealization providing a better fit. It seems likely that these observations are a consequence of the Aboav-Weaire law reflecting dominantly topological characteristics, whereas Lewis's law depends more on the specific character of the inter-vertex forces and constraints.

It would be interesting to try to extend the data set for both experimental and computer samples of these Zachariasen-like amorphous networks, by varying their preparation conditions.

5. **Acknowledgements** We should like to thank Mike Treacy and Walter Whiteley for continuing useful discussions, and P. Huang for providing the additional unpublished data from the Cornell group. AK would like to acknowledge funding from GAANN P200A090123 and the ARCS Foundation. MFT would like to thank Theoretical Physics at the University of Oxford for continuing hospitality and the National Science Foundation for support under grant DMR 0703973.